# An Investigation into the Mathematical Nature of Electrophysiological Signals with Applications

Kaushik K. Majumdar, Anagh Pathak, Viswadeep Sarangi

*Abstract*—In this work we have proposed, possibly for the first time, a rigorous mathematical definition for the one dimensional time domain electrophysiological signals and established its relationship with two of the three Dirichlet's conditions. We have argued that any such signal can be represented as the trajectory of a particle moving in a force field with one degree of freedom. At each point on the trajectory, that is, on the signal, the kinetic energy dissipated by the particle embeds semantic information into the trajectory or the signal in terms of giving its shape. We have shown that the rate of kinetic energy dissipation operator or the power operator P is of importance in shape analysis of the signal by considering its sign changes. Operating the P-operator on digital signals we have mathematically proved that its sign change can induce 13 different shapes to a three successive point configuration. We have shown that the entropy of distribution of these 13 different shapes or configurations or features across focal intracranial electroencephalogram (iEEG) signals from patients with epilepsy can distinguish the epoch before an epileptic seizure from the epoch during the seizure in a statistically significant way. It has also been shown that these 13 features can clearly distinguish between raw signals, their shifted surrogates, the power spectrum preserved shifted surrogates and Gaussian white noise signals, provided signals of a minimum length are available.

## I. INTRODUCTION

THE term *signal* is generally applied to something that conveys information [1], [2]. It is reasonable to visualize signal as a function of space (such as, an image) or a function of time (such as, an electrocardiogram) or of both [1]. A signal may consist of other attributes as well, for example, discrete symbols alone, such as, telegraphic signal. However, a precise mathematical specification for a function to be a signal is absent in the literature. In context of validity of certain mathematical operations, say, Fourier transform, specific conditions on the function are imposed [3]. Same is the case for wavelet transforms [4].

Biomedical signals on the other hand are a special class of signals originating from the living beings. Initially those signals used to be largely collected from humans only for clinical purposes. But nowadays a substantial portion of this acquisition is for research purpose and also from nonhuman animals. Often we do not have any control over the sources and in many instances the signal generation itself is poorly understood, if at all. Also traditionally, biomedical signals have been processed by visual inspection by clinicians and experimentalists, which still has remained the gold standard for applications like seizure detection [5]. So, apart from accommodating validity of mathematical operations associated with common signal processing algorithms, biomedical signals will have to be *plotable*, that is, it must be possible to generate the time versus amplitude graph of the signal. We will see that the function representing a signal must have some special properties to be plotable in the above sense. Special purpose software like EEGLAB [6], SPM [7], FSL [8], etc. are widely used for plotting and visualizing particular biomedical signals.

Electrophysiological signals are a subset of biomedical signals, which are due to electric potential of physiological origin varying over the time. Electrophysiological signals are typically collected placing electrodes in appropriate locations of the body. Each electrode collects a one dimensional time domain signal, whose amplitude is the potential difference between the electrode and a reference electrode. The exact origin of the electrophysiological signals may be ambiguous for some types of signals [9]. For some other types it may be understood better [10]. However, the primary source of physiological currents is the membrane current due to the ion channels activities. Any membrane voltage has two states, *depolarization* (increasing) and *hyperpolarization* (decreasing) which shapes the time versus voltage signal. Sequences of depolarization and hyperpolarization and their rate with respect to time determine the shape of the membrane voltage signal. Any electrophysiological signal (possibly among others) contains superposition of membrane voltage signals. So, the shape of any electrophysiological signal is affected by membrane depolarization and hyperpolarization within a neighborhood. We will see that the P-operator can identify an exhaustive list of thirteen different shape patterns in a discrete signal segment consisting of three points. It will be shown that in most cases the entropy of distribution of these shapes in

K. K. Majumdar, is with the Systems Science and Informatics Unit, Indian Statistical Institute, 8th Mile, Mysore Road, Bangalore 560059, India (e-mail: mkkaushik@hotmail.com).
A. Pathak, is with the Systems Science and Informatics Unit, Indian Statistical Institute, 8th Mile, Mysore Road, Bangalore 560059, India (e-mail: pathak02@gmail.com).
V. Sarangi, is with the Systems Science and Informatics Unit, Indian Statistical Institute, 8th Mile, Mysore Road, Bangalore 560059, India (e-mail: deepun.deep@gmail.com).

focal intracranial EEG (iEEG) channels goes down prior to epileptic seizures. We will also see treating these thirteen shapes as conventional features can very clearly distinguish between the raw signals, their shifted surrogates and the power spectrum preserving shifted surrogates. For this, signal segments above a critical duration are needed.

In the next section we present a mathematical formalization of analog electrophysiological signals and we will also establish that by P-operator an exhaustive list of thirteen different shapes in a discrete one dimensional time domain signal segment consisting of three points can be identified. In section 3 we will study the entropy of distribution of the thirteen different shapes in focal iEEG signals of patients with epilepsy before and during seizures. In section 4 with the help of the thirteen features we will separate out original signals, their shifted surrogates and power spectrum preserving shifted surrogates from a mixture of all the three types. The paper concludes with a Conclusion section, which also incorporates future directions.

## II. MATHEMATICAL FORMALISM

### A. Electrophysiological Signal

A number of important electrophysiological signals are generated in the nervous systems, for example, neuronal spike train [11], local filed potential (LFP) [11], electrocorticogram (ECoG) [11] and electroencephalogram (EEG) [11]. Apart from the neuronal spike trains the debates surrounding the origin of the other three have not been completely settled down. Exactly how different ionic potentials, morphological shape of the membranes and tissue impedances contribute in them has not been fully understood yet.

An electrophysiological signal $s(t)$ originates due to simultaneous potential difference build up across membranes and their summation [9], [10]. As $s(t)$ propagates from its source to a sensing electrode, it may be modified by ionic current contribution, local geometric configuration of the medium, tissue impedance, network properties, presence of neurotransmitters, sensory electrode configuration and other factors [11], [12]. Nevertheless, $s(t)$ can be thought as the trajectory of a particle moving in a time varying force field, no matter however complicated, with one degree of freedom for the movement (along amplitude, that is, perpendicular to the time axis). With this in mind, classical particle dynamical system theoretic tools may be brought into the domain of processing $s(t)$ as a signal.

As $s(t)$ has been formed by the motion of a particle, its shape has been endowed by the nonlinear motion of the particle. At a time $t$ the acceleration of the particle is $s''(t) = \frac{d^2 s(t)}{dt^2}$, which is also the force acting on the particle at that time assuming its mass is unit. Work done by the particle for an infinitesimal displacement $ds$ at that instant is $\frac{d^2 s(t)}{dt^2} ds(t)$, which takes time $dt$, and therefore the rate of work done is

$$P(s(t)) = \frac{d^2 s(t)}{dt^2} \frac{ds(t)}{dt}, \qquad (1)$$

which is the power of the particle in sense of the particle dynamics. $P(s(t))$ is the rate at which the kinetic energy of the particle is giving shape to its trajectory, that is, the signal $s(t)$ in order to embed 'information' into the signal. This information (not in the sense of Shannon) may be useful or useless (noise) depending on the objective of the user. However, this particle trajectory analogy holds for any one dimensional, time domain analog signal, and therefore $P(s(t))$ is the rate at which information (not in sense of Shannon, but the actual semantic or meaning of the signal) is being embedded or etched or written into the signal.

*Definition 1*: An analog *electrophysiological signal* $s$ is a function $s : \mathfrak{R}^+ \to \mathfrak{R}$, such that,
(1) $s$ is piecewise continuous,
(2) $s$ is bounded on a compact subset of $\mathfrak{R}^+$, the set of nonnegative real numbers, where $\mathfrak{R}$ is the set of all real numbers.
(3) $s$ may be nondifferentiable at most at a finitely many points of a compact subset of $\mathfrak{R}^+$.

*Definition 2*: A signal $s(t)$, $t \in [a, b]$ is *plotable* if and only if the graph of the signal $(t, s(t))$ can be traced along the tangent at each point on the graph.

According to Definition 2, everywhere continuous, nowhere differentiable functions (Theorem 7.18 of [13], p. 154) are not plotable, because the tangent does not exist at any point on the graph. In this sense most of the continuous functions are not plotable. The analog paper EEG and ECG are plotable signals. The ink pen or marker that is used to generate the paper EEG or paper ECG always traverses the curve along the tangent at each point.

*Lemma 1*: Condition (3) in Definition 1 implies that the signal is plotable.

*Proof*: Condition (3) implies that for any nondifferntiable point $t = a$, there will either be no other nondifferntiable point or there will be one $t = b$, such that, $s'(b)$ does not exist and $s'(t)$ exists for $a < t < b$ or $b < t < a$ as the case may be. This means, it is possible to draw tangent at any point on the curve $(t, s(t))$ for $t$ between $a$ and $b$ (or on the left side of $a$ and on the right side of $a$, when $a$ is the only point of nondifferentiability). This completes the proof.

*Lemma 2*: If there are infinitely many points of nondifferentiability of $s(t)$ on a compact subset $\square$ of $\mathfrak{R}^+$ then the signal curve will not be plotable in a maximal open subset $\Omega \subset \square \subset \mathfrak{R}^+$.

*Theorem 1*: Every sequentially compact metric space is compact and vice versa.

*Proof*: Please see p. 124 of ref. [14].

*Proof of Lemma 2*: Theorem 1 implies, if $\{x_n\}_{n=1}^{\infty} \subset \mathbb{R}$, such that $s'(x_n)$ does not exist for any $n$, then there must be $x \in \mathbb{R}$ and $\lim_{n' \to \infty} x_{n'} = x$ ($n'$ takes values from a subset of natural numbers), so that $s'(x)$ does not exist. There will be $\delta > 0$ neighborhood $N_\delta(x)$ of $x$ for any $\delta$ which will contain an infinite convergent subsequence of $\{x_n\}$ on which $s'$ does not exist. This is true for $\delta \to 0$ as well. Let $y$ is on the boundary of $N_\delta(x)$, that is, $y + \delta = x$. Consider $\frac{s(y+h)-s(y)}{h}$, where $h = \delta$. $\delta \to 0$ implies $h \to 0$ and

$$s'(y) = \lim_{h \to 0} \frac{s(y+h)-s(y)}{h} = \lim_{h \to 0} \frac{s(x)-s(x-h)}{h} = s'(x)$$

Since $s'(x)$ does not exist, $s'(y)$ also does not exist. But $y \in N_{2\delta}(x)$ for $\delta > 0$. Repeating the above logic for any point $z \in N_{2\delta}(x)$ and considering $z - y = \alpha h$ for $0 < \alpha < 1$, and

$$s'(z) = \lim_{\alpha h \to 0} \frac{s(z-\alpha h)-s(z)}{\alpha h} = \lim_{\alpha h \to 0} \frac{s(y)-s(y+\alpha h)}{\alpha h},$$

we conclude that $s'$ does not exist on any point of $N_{2\delta}(x)$. We take the union of all such open neighborhoods within $\mathbb{R}$ and denote it by $\Omega$. In other words, if there are infinitely many points in $\mathbb{R}$, on which $s'$ does not exist, then there will be an entire interval of nonzero length on which $s(t)$ will not be plotable. A connected open component of $\Omega$ will be one such interval of $\mathbb{R}^+$. This completes the proof.

*Theorem 2*: Analog electrophysiological signals are plotable over a closed and bounded time interval, except possibly at a finite number of jump discontinuities.

*Proof*: By Heine-Borel theorem (Theorem G of Chapter 4 of ref. [13]), a closed and bounded interval of $\mathbb{R}^+$ is compact. By condition (2) of Definition 1, $s$ is then bounded on any closed and bounded interval of $\mathbb{R}^+$. Therefore, any discontinuity of $s$ on that interval will be a jump discontinuity. Only a finite number of such discontinuities are possible in a closed and bounded interval, by the Heine-Borel theorem and condition (3) of Definition 1 (nondifferentiable points include points of discontinuity as well). By Lemma 1 the analog signal $s$ is plotable in the interval between any two successive jump discontinuities. This completes the proof.

It is worth noting that Definition 1 guarantees that any analog electrophysiological signal satisfies two of the three Dirichlet's conditions (see p. 237 of ref. [3]). On a closed and bounded time interval $s(t)$ has only finite number of points of nondifferentiability implies that $s$ has only a finite number of points of discontinuity on that interval. Since $s$ is bounded on such intervals, it only has finite number of jump or finite discontinuities. Next, we will show that $s$ has finite number of maxima and minima on such intervals. We will show it by contradiction. Let it is not. That is, there exists a closed bounded interval of $I \subset \mathbb{R}^+$, such that, $s$ attains infinite number of maxima on $I$. We subdivide $I$ into closed and bounded intervals $I_1$ and $I_2$, such that $I = I_1 \cup I_2$, $I_1 \cap I_2 = \{c\}$, where $c$ is the boundary point between $I_1$ and $I_2$, and $|I_1| = |I_2|$. Let $s$ has infinite number of maxima in $I_1$. Bisect $I_1$ exactly the same way. If we keep iterating it, we will get a closed and bounded interval $I_k$, such that $|I_k| \to 0$ and yet $I_k$ contains infinite number of points of maxima of $s$. Let $m$ be one such point of maximum. Then $s(m - h_1)$ is increasing for $h_1 \to 0$, $s(m + h_2)$ is decreasing for $h_2 \to 0$, where $h_1 > 0$ and $h_2 > 0$. Since there are infinite such $m \in I_k$ and $|I_k| = |I|2^{-l}$ if $I_k$ is obtained at the $l$th iteration, $\frac{s(m)-s(m-h)}{h}$, $0 < h < h_1 < |I_k|$ will be a positive number (as $s$ is increasing in $(m - h_1, m)$. The positive number, and will keep increasing as $l$ increases. Similarly, $\frac{s(m+h)-s(m)}{h}$, $0 < h < h_2 < |I_k|$ will be a negative number, and will keep decreasing as $l$ increases. That is, $\lim_{h \to 0} \frac{s(m)-s(m-h)}{h}$ and $\lim_{h \to 0} \frac{s(m+h)-s(m)}{h}$ will never be equal as $l$ will increase arbitrarily. In other words, $s'(m)$ will not exist. Since there are infinite such $m \in I_k \subset I$, $s$ will not be plotable by Lemma 2. This contradiction implies that there must be a finite number of maxima of $s$ in any closed and bounded interval of $\mathbb{R}^+$. Similarly, it can be shown that there will be only a finite number of minima of $s$ in any such interval.

However, it can be shown by a counter example (which will be a bit tedious) that satisfying Dirichlet's conditions will not make a continuous function an analog electrophysilogical signal according to Definition 1.

### B. Digital Signals

A discrete or digital time domain signal can still be modeled as the trajectory of a particle moving in a force field with only one degree of freedom, but with as many points of nondifferentiability in any one second long segment as the sample frequency. Moreover, the slope between any two successive points will remain constant. Clearly, in case of the digital signals most of the above notions and arguments cannot be carried forward. However, the discrete version of power operator or P-operator (P( ) as given by equation (1)) can still give us interesting insights into the digital signals. We will denote a digital signal by $x[n]$, where $n$ is a nonnegative integer.

The P-operator is a product between the first derivative and the second derivative. But there are subtle intricacies that go into the implementation of the operator on a digital signal. If the original signal is represented by $x[n]$, the first difference as $x'[n]$, second difference as $x''[n]$ and the P-operated discrete signal as $P(x[n])$, then each point in $P(x[n])$ can be calculated using the points $x[n-1]$, $x[n]$ and $x[n+1]$ in the signal, where n can vary from $1$ to $N$, the number of discrete samples in the signal. The first difference is calculated using a backward difference $x'[n] = x[n] - x[n-1]$. The second difference is calculated as $x''[n] = x[n+1] - 2x[n] + x[n-1]$. We are interested in the change of the sign of the product $x''[n]x'[n]$, which we will call *left product* and $x''[n]x'[n+1]$, which we will call *right product*. It has been elaborated in Fig 1.

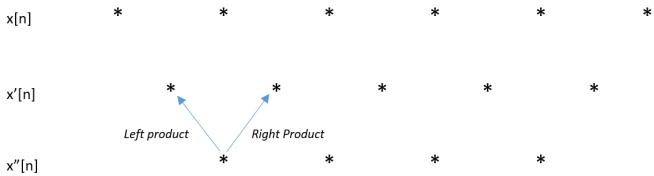

Fig 1. Left product and right product. For successive points n – 1, n and n + 1, the left product is denoted by P(x[n-]) and the right product is denoted by P(x[n+]).

It is easy to see that, if
$$P(x[n-]) < 0 \text{ and } P(x[n+]) > 0 \quad (2)$$
hold, then $x[n]$ is a local optimum. If in addition to (2)
$$x''[n] < 0, \quad (3)$$
also holds, then $x[n]$ is a local maximum or a peak. Similarly, if in addition to (2),
$$x''[n] > 0, \quad (4)$$
also holds, then $x[n]$ is a local minimum or a trough (see ref. [15] for detailed derivation). This is advantageous, because in digital signals first order difference does not vanish at local optima. Moreover, it helps us detect all peaks and troughs in the digital signal irrespective of their shape and size. We have studied simultaneous occurrence of such peaks and troughs across focal and nonfocal channels before, during and after epileptic seizures in [15].

Notice that, $P(x[n-]) = x''[n]x'[n]$ and $P(x[n+]) = x''[n]x'[n+1]$. In both the factors of $P(x[n-])$ and $P(x[n+])$ $x''[n]$ is common, and therefore there are only three factors in total in $P(x[n-])$ and $P(x[n+])$, namely $x'[n]$, $x''[n]$ and $x'[n-1]$. Each of these three factors can either be positive or negative or zero. Therefore, sign change in P-operator can happen in $3^3 = 27$ ways. It is worth noting,

*Lemma 3*: $P(x[n-]) > 0$ and $P(x[n+]) < 0$ is impossible to take place.

*Proof*: Let there are three points $n-1$, $n$ and $n+1$, for which $P(x[n-]) > 0$ and $P(x[n+]) < 0$ hold. The following two cases are all that are possible.
Case 1: $x''[n] > 0$ and $x'[n] > 0$, and $x''[n] > 0$ and $x'[n+1] < 0$. Since $x'[n] > 0$ and $x'[n+1] < 0$, $x[n] > x[n-1]$ and $x[n+1] < x[n]$. That is, $2x[n] > x[n-1] + x[n+1]$. But, $x''[n] = x[n+1] - 2x[n] + x[n-1] > 0$, which implies $x[n+1] + x[n-1] > 2x[n]$, leading to a contradiction.
Case 2: $x''[n] < 0$ and $x'[n] < 0$, and $x''[n] < 0$ and $x'[n+1] > 0$. Similar argument as in Case 1 will lead to a contradiction for this case also.
Therefore $P(x[n-]) > 0$ and $P(x[n+]) < 0$ is an impossibility. This completes the proof.

In the previous subsection we have seen that P-operator operated on a signal gives the rate at which information (not in sense of Shannon, but in terms of the semantics of the signal as carrier of signatures of undergoing physiological processes) is being embedded into the signal in the form of creation of its shape. When P-operator goes from negative to positive in a discrete signal it signifies the most significant transformation of kinetic energy of the hypothetical moving particle (whose trajectory the signal is) into information to be embedded into the signal (the other possibilities are 0 to +, + to + and – to – (0 to – and + to 0 are not possible by Lemma 6), none of which is as much a transition as from – to + as the sequence of signs is –, 0, +). Also when P-operator changes sign from – to + the signal must have either a peak or a trough. Putting these two together we observe that peaks and troughs are the most information rich events of an electrophysiological signal, which is supported by the Hodgkin-Huxley model of action potential generation [16] and PQRST model of ECG spikes [17]. Since P-operator cannot go from positive to negative (Lemma 3), information embedded into the signal cannot be transformed back to the kinetic energy of the hypothetical particle whose trajectory is the signal.

*Lemma 4*: It is impossible to hold (1) $x"[n] = 0$ and $x'[n] > 0$ and $x'[n+1] < 0$, or (2) $x"[n] = 0$ and $x'[n] < 0$ and $x'[n+1] > 0$.

*Proof*: $x"[n] = 0$ implies $2x[n] = x[n-1] + x[n+1]$. $x'[n] > 0$ and $x'[n+1] < 0$ together imply $2x[n] > x[n-1] + x[n+1]$, which is contradictory to $2x[n] = x[n-1] + x[n+1]$. So (1) is impossible. Similarly it can be established that (2) is also impossible. This completes the proof.

*Lemma 5*: It is impossible (1) $x"[n] > 0$ and $x'[n] = 0$ and $x'[n+1] = 0$, or (2) $x"[n] < 0$ and $x'[n] = 0$ and $x'[n+1] = 0$.

*Proof*: As in Lemma 4.

*Lemma 6*: Transition of P-operator from 0 to – and + to 0 are not possible.

*Proof*: $P(x[n-]) = 0 \Rightarrow x"[n]x'[n] = 0 \Rightarrow x'[n] = 0$ ($x"[n] \neq 0$ as $P(x[n+]) \neq 0$) $\Rightarrow x[n] = x[n-1]$. Since $P(x[n+]) < 0$,
$(x[n+1] - 2x[n] + x[n-1])(x[n+1] - x[n]) < 0 \Rightarrow$
$(x[n+1] - x[n])(x[n+1] - x[n]) < 0$ (using $x[n] = x[n-1]$) $\Rightarrow (x[n+1] - x[n])^2 < 0$, which is impossible. Arguing similarly it can be shown that transition of P-operator from + to 0 is also impossible. This completes the proof.

*Lemma 7*: Transition of P-operator from 0 to 0 is possible only in the following three ways:
1. $x'[n] > 0$, $x"[n] = 0$, $x'[n+1] > 0$,
2. $x'[n] < 0$, $x"[n] = 0$, $x'[n+1] < 0$,
3. $x'[n] = 0$, $x"[n] = 0$, $x'[n+1] = 0$.

*Proof*: 0 to 0 transition of P-operator can happen in 11 different ways. It is easy to check that the cases 1, 2, 3 above are all possible to occur. If $x"[n] \neq 0$, both $x'[n]$ and $x'[n+1]$ must vanish. This can happen in two different ways and both can be ruled out by Lemma 5. Besides, when $x"[n] = 0$, none of (1) $x'[n] < 0$ and $x'[n+1] > 0$, and (2) $x'[n] > 0$ and $x'[n+1] < 0$ is possible by Lemma 4. This implies 4 more cases are left to be investigated. These cases are as following:
(a) $x'[n] > 0$, $x"[n] = 0$, $x'[n+1] = 0$,
(b) $x'[n] < 0$, $x"[n] = 0$, $x'[n+1] = 0$,
(c) $x'[n] = 0$, $x"[n] = 0$, $x'[n+1] > 0$,
(d) $x'[n] = 0$, $x"[n] = 0$, $x'[n+1] < 0$.

The proof of each of (a), (b), (c) and (d) is similar to the proof of Lemma 4 and Lemma 5. This completes the proof.

Lemmas 3, 4 and 5 together rule out 6 possibilities out of the total 27, that the sign change of P-operator introduces into a digital signal. Lemma 6 rules out 4 more possibilities, because for each of the transitions from 0 to – and from + to 0, $x"[n]$ can be either positive or negative. Lemma 7 rules out another four possibilities. So, there are a total of 14 possibilities out of 27 which can be mathematically ruled out. We have essentially proved the following.

*Theorem 3*: Sign change of P-operator can give rise to precisely 13 different configurations for any 3 successive points in a digital signal as furnished in Table I.

We have shown that $P(x[n])$ is the rate, at which the kinetic energy of the hypothetical moving particle traversing $x[n]$, is being converted into semantic information to be embedded into the shape or form of $x[n]$. It is clear that this is the only way semantic information can be embedded into a one dimensional time domain signal. From this point of view Theorem 3 asserts that semantic information can be embedded into a one dimensional time domain digital signal in precisely 13 different 3-point configurations as illustrated in Table I.

*Semantic information* pertains to the meaning or the interpretation of the signal and is different from Shannon's notion of information. We can illustrate it by a fair coin toss. Before the toss the outcome is uncertain with probability ½ for either head or tail. So the entropy of the event is 1 if the base of the logarithm is 2, which is also the Shannon information content of the event. Here, the semantic information is 0, because it is yet to be decided and therefore is not available for the moment. After the toss, the Shannon information content becomes 0 and the semantic information becomes 1, because head or tail is representable by 1 bit.

TABLE I
Analytical and geometrical forms of all the 13 configurations possible by the sign change of P-operator on discrete signals

| SIGN CHANGE | $x'[n]$ | $x''[n]$ | $x'[n+1]$ | PATTERN |
|---|---|---|---|---|
| ++ | + | + | + | |
|    | − | − | − | |
| −− | + | − | + | |
|    | − | + | − | |
| −+ | − | + | + | |
|    | + | − | − | |
| 00 | + | 0 | + | |
|    | − | 0 | − | |
|    | 0 | 0 | 0 | |
| 0+ | 0 | + | + | |
|    | 0 | − | − | |
| −0 | + | − | 0 | |
|    | − | + | 0 | |

Different patterns formed in discrete electrophysiological signals have been studied even with physical significances [18]. In P-operator up to second degree difference is involved and therefore it can determine the shape of a signal segment consisting of three successive points. In [18] it has been shown that a $n$ point long discrete signal segment can form $n!$ different patterns based on how saw-tooth like they appear with different tooth sizes and angular inclination between two successive edges. Each permutation of the teeth will give one particular pattern. Roughly speaking, here in this work the 13 configuration 3 point patterns are offering 7 additional configurations based on not only the size of the tooth, but also on different degree of sharpness of the subtended angle between any two successive edges.

## III. APPLICATION TO NEURAL SIGNALS

### A. Data

ECoG data of 21 epileptic patients containing 87 focal onset seizures have been obtained from the Freiburg Seizure Prediction Project (https://epilepsy.uni-freiburg.de/2008). One hour recording containing preictal, ictal and postictal ECoG of 1 h duration in each of the 87 cases is available. The ECoG data were acquired using Neurofile NT digital video EEG system (It-med, Usingen, Germany) with 128 channels, 256 Hz sampling rate, and a 16 bit analog to digital converter. In all cases the ECoG from only six sites have been analyzed, because only six channel data were made available through the above website. However this is a publicly available data set and therefore good for benchmarking novel algorithms. Three of the six channel data are from the focal areas and the other three from outside the focal areas. For each patient there are 2–5 h of ictal data (actually preictal + ictal + postictal) recordings. Each hour's recording contains only one seizure of few tens of seconds to a couple of minute duration [19].

### B. Results

It is well known that entropy of the iEEG signal goes down during an epileptic seizure [20]. However, estimating entropy of a signal is a tricky issue [21]. The easiest and perhaps the most popular method of estimating entropy of a signal is by histogram, where the bin size is arbitrary and affects the result significantly [21]. Here we have only 13 different features (Theorem 3) and therefore within a signal window we can have only 13 histograms, each containing the number of times a particular feature occurs in that window. We eventually get a frequency distribution of the features in that window and then keep sliding the window in non-overlapping manner to cover the given signal segment. This way we have measured the Shannon entropy in a focal channel during seizure (with a 1s, i.e. 256 point window) and before the seizure onset and after the seizure offset (onset and offset points were demarcated by certified epileptologists at the source clinic) for a duration equal to that of the seizure duration. In a predominant majority of the cases entropy during seizure went down compared to before seizure in a statistically significant way as can be seen in Table II. Although there are 87 seizures, for 2 of them signal of sufficient length was not available prior to seizure to carry out the study.

TABLE II
Summary of patient specific result of difference in Shannon entropy based on the 13 features in Theorem 3. Statistical significance has been determined by Kruskal-Wallis test [23].

| Patient | Mean duration(s) | Seizures | Seizures with P<0.05 | Seizures with P> 0.05 |
|---|---|---|---|---|
| 1 | 13.1 | 4 | 1 | 3 |
| 2 | 118.2 | 3 | 3 | 0 |
| 3 | 92.7 | 5 | 4 | 1 |
| 4 | 87.4 | 5 | 4 | 1 |
| 5 | 44.9 | 5 | 2 | 3 |
| 6 | 66.9 | 3 | 2 | 1 |
| 7 | 153.5 | 3 | 3 | 0 |
| 8 | 163.7 | 2 | 0 | 2 |
| 9 | 113.7 | 4 | 4 | 0 |
| 10 | 411 | 5 | 4 | 1 |
| 11 | 157.3 | 4 | 3 | 1 |
| 12 | 55.1 | 3 | 3 | 0 |
| 13 | 158.3 | 2 | 1 | 1 |
| 14 | 216.4 | 4 | 3 | 1 |
| 15 | 145.4 | 4 | 3 | 1 |
| 16 | 121 | 5 | 5 | 0 |
| 17 | 86.2 | 5 | 4 | 1 |
| 18 | 13.7 | 5 | 1 | 4 |
| 19 | 12.5 | 4 | 3 | 1 |
| 20 | 85.7 | 5 | 5 | 0 |
| 21 | 83.1 | 5 | 4 | 1 |
| Total | | 85 | 62 | 23 |

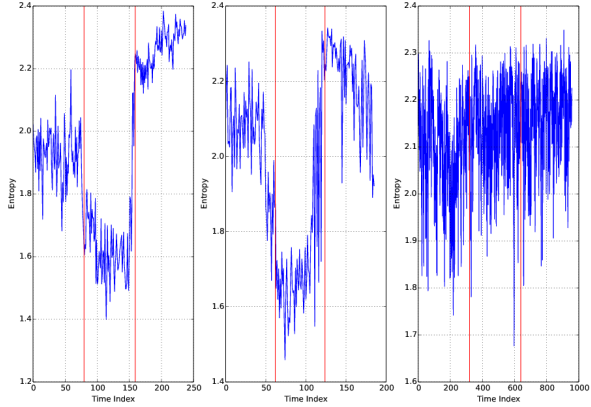

Fig 2. Time index versus entropy plot for all three seizures of patient 7. Red vertical lines indicate seizure onset and offset determined by certified epileptologists. Entropy is clearly lower during the first two seizures. It also went down during the third with p < 0.001, although that is visually less obvious.

The trend between the entropy during and after the seizure is not very clear across the population. This can be due to the fact that after the seizure the entropy remains low for some time, because the iEEG signal across the focal channels after the seizure offset almost dies down. Frequency distribution of the 13 features of Theorem 3 (Table I) is not normal in general and therefore we went for Krsukal-Wallis test (p. 198 – 201 of ref. [22]) to measure the statistical significance of the difference in entropy between before and during the seizure. For each seizure we have generated plot like Fig 2. It is clear from the plot that the entropy during the seizure is less than prior to its onset.

## IV. RANDOM SIGNALS

### A. Data

The CPGJNM2012 dataset is a collection of 95 simulations of 10 minutes long extracellular potential data with a sampling rate of 24 KHz. The dataset was created and validated in the Department of Engineering, University of Leicester, UK [23]. The version we used was made publicly available in 2012. It can be freely obtained from the URL: http://bioweb.me/CPGJNM2012-dataset

### B. Results

To study yet another application of the 13 configurations of Theorem 3 (Table I) derived from the P-operator, we have generated three different datasets out of the CPGJNM2012 dataset namely, the original signals $x[n]$s, shifted surrogate signals $R(x[n])$s and the shifted surrogate signals preserving the power spectrum density $Q(x[n])$s.

The values of the original signal $x[n]$ were randomly shuffled across the time to generate the shifted surrogate signal $R(x[n])$. To generate the shifted surrogate signal by preserving the power spectrum density across the frequencies, the following technique was utilized. The fast Fourier transform (FFT) of the original signal was obtained, and the resulting complex number array was split into two parts, the amplitude of the complex number and the phase. Keeping the amplitudes in the same place, the phase values were randomly shuffled across the amplitudes of the signal. The resulting duplet signal was multiplied using $r \exp(i\theta)$ convention, where $r$ represents the amplitude part of the signal, $\theta$ represents the new phase of the corresponding amplitude after shuffling. An inverse fast Fourier transform (iFFT) was performed on the resulting complex number signal to obtain the shifted surrogate signal while preserving the power spectrum density $Q(x[n])$.

The P-operation was performed on four types of signals $x[n]$ (from the CPGJNM2012 dataset), $R(x[n])$, $Q(x[n])$ and Gaussian white noise signal $G[n]$ (generated in MATLAB with same length and sample frequency as $x[n]$) to obtain the 13 configurations' frequency distribution decomposition of the signals within a window. Shifting the window in non-overlapping manner we got as many 13 dimensional vectors as there are windows in the whole signal segment. These 13 dimensional vectors formed four clusters, one each for the types $x[n]$, $R(x[n])$, $Q(x[n])$ and $G[n]$ respectively. Clustering was performed in 13 dimensional space to underscore the efficacy of the configurations or features as input to a unsupervised classification algorithm. The results show a clear distinction between the clusters formed with very high inter-cluster distance and very low distance between any pair of points within a single cluster. The top subplot of Fig 3 shows the plot of clustering for 3 maximum variance (of the frequency values of a configuration across the signals) configurations out of 13. The bottom plot of Fig 3 is a two dimensional projection of the three dimensional plot to get the optimum separation among the clusters.

signal length, then 30% and so on. For each of these lengths we determined the maximum Euclidean distance between any two 13 dimensional points from shifted surrogate signals. Fig 4 shows the maximum Euclidean distance between any two 13 dimensional points versus the window length plot. Notice that the plot becomes almost stable at the 60% of the original signal length (at 6 minutes of the 10 minutes long signals) window length. For the power spectrum preserving shifted surrogate signals the plot becomes stable at the 30% of the original signal length (at 3 minutes of the 10 minutes long signal). Similarly, for the raw signal it is 80% of the signal length, that is, 8 minutes. The shifted surrogates of CPGJNM2012 are closely similar to Gaussian white noise signals, but they are not identical (Fig 3), whereas the shifted surrogates of the Freiburg signals are much further apart from the Gaussian white noise signals (Fig 5).

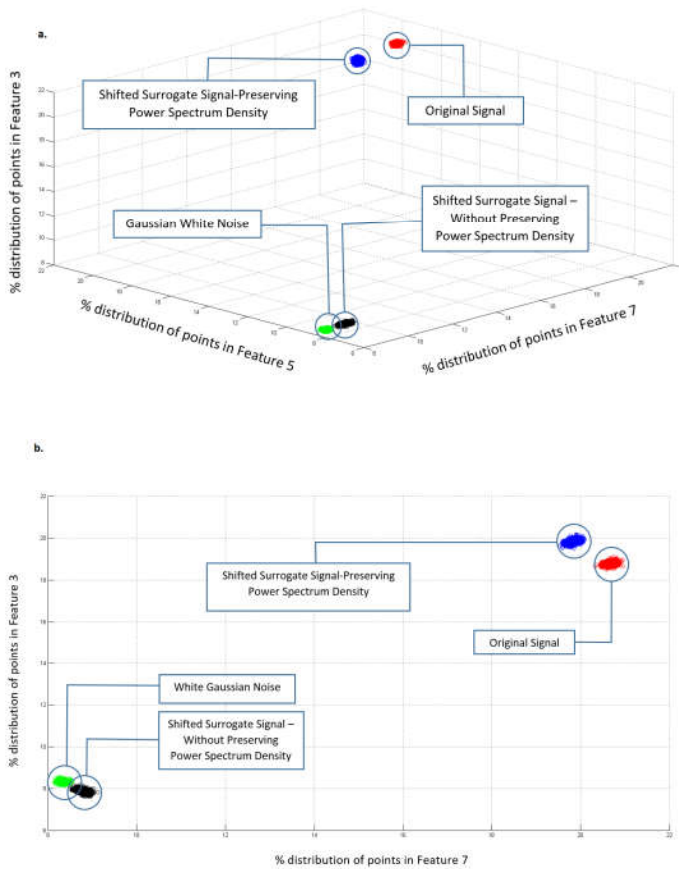

Fig 3. Plots of best 3D (a) and 2D (b) separation of the CPGJNM2012 dataset from its shifted surrogate, power spectrum preserving shifted surrogate and Gaussian white noise signals.

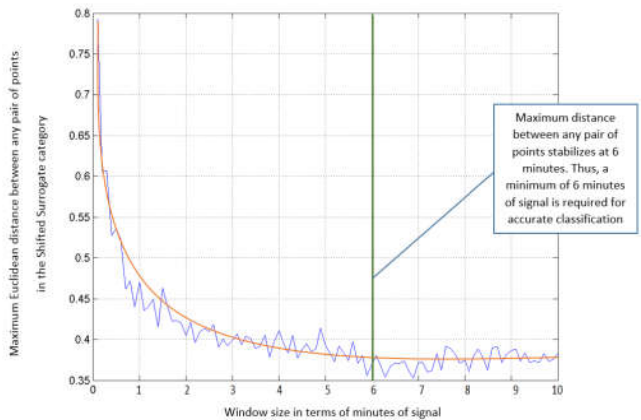

Fig 4. The plot of the signal window length versus maximum distance between any two points in the 13 dimensional feature or configuration space for the shifted surrogate signals of CPGJNM2012. The smooth curve fitting through the two dimensional points has also been shown.

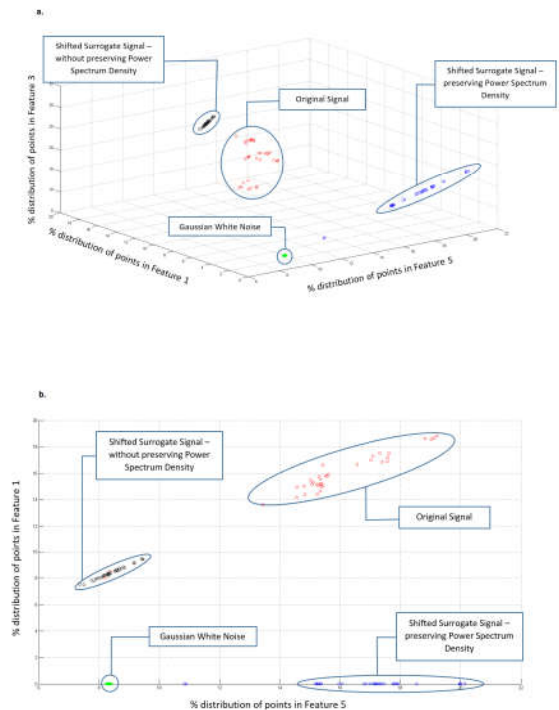

Fig 5. Same as in Fig 3, but on the Freiburg iEEG data. Here the signals are only 9,21,600 time points long, whereas in Fig 3 the length was 144,00,000 time points.

For this optimum separation of the clusters a minimum window length is required for raw signals, shifted surrogates and power-spectrum preserving shifted surrogate signals (separate length for each category). In Fig 4 we have presented the optimum window length by simulation for the shifted surrogate signals of CPGJNM2012. Initial window length was 10% of the whole signal length. Then it was 20% of the whole

It is clear from Fig 5 that the analysis for the Freiburg data gives a bit different results. Apart from the larger distinction between the shifted surrogate and the Gaussian white noise, the shifted surrogate itself could not be separated as clearly from the raw signal as it was possible for the CPGJNM2012 data. A minute inspection of the colored version of Fig 5 will reveal that some 3 dimensional (out of 13) points belonging to the raw signals falling inside the cluster of the points belonging to the shifted surrogate signals, which is more clear in Fig 5(b) than in Fig 5(a). Also the clustering of the raw Freiburg data is not as tight as the raw CPGJNM2012 data. The signal length in the CPGJNM2012 data was 144,00,000, whereas the signal length in the Freiburg data was 9,21,600. The optimum window size for clustering the raw Freiburg signals is again 80% of the signal length (Fig 6), that is, 48 minutes of the 60 minutes long signals. For the shifted surrogate with PSD preserved it is again 60% of the signal length, that is, 36 minutes. For the shifted surrogate signals it is 37% of the window length or 22 minutes. For the Gaussian white noise signals (generated in MATLAB) 50% of the total signal length as the window length will give the tightest clustering as can be seen from Fig 7.

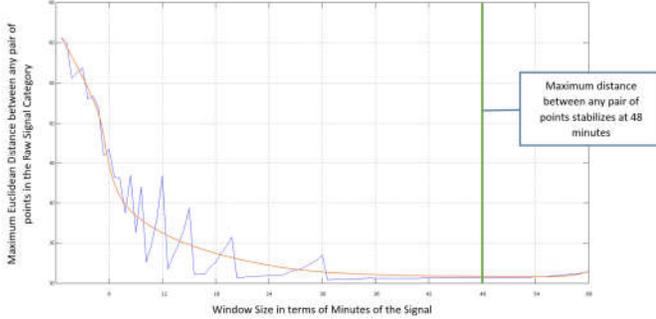

Fig 6. The plot of the signal window length versus maximum distance between any two points in the 13 dimensional feature space for the raw signals of the Freiburg dataset. The smooth curve fitting through the two dimensional points has also been shown.

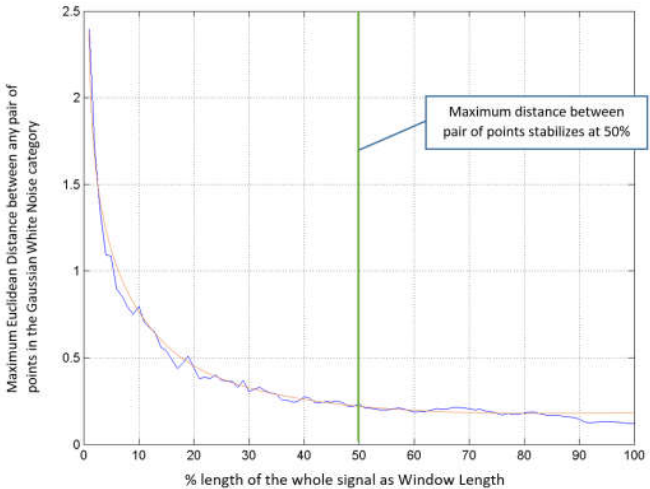

Fig 7. The plot of the signal window length versus maximum distance between any two points in the 13 dimensional feature space for the Gaussian white noise signals. Here the number of time points is 9,21,600. The smooth curve fitting through the two dimensional points has also been shown.

## V. Conclusion

Signals are of predominant importance to engineers from certain disciplines. But signal is processed by Mathematics, Statistics and Computation. Having a mathematical definition of signals will be advantageous. In this work we have proposed a rigorous mathematical definition for one dimensional, time domain, analog electrophysiological signals, which can be readily extended to any one dimensional, time domain, analog signals. We have proved this definition implies two of the three Dirichlet's conditions. The third one $\int_{-\infty}^{\infty} |s(t)| dt < \infty$ is satisfied by most of the real life signals. So, Definition 1 subsumes Dirichlet's conditions, and at the same time guarantees that the electrophysiological signals can be plotted like paper EEG and ECG, which is essential for visual inspection by the clinicians and the experimentalists.

We have visualized any analog, one dimensional time domain signal as the trajectory of a particle moving in a force field with one degree of freedom. From this point of view we introduced the power operator or the P-operator, which gives the rate at which the kinetic energy of the (hypothetical) particle is being transformed into the semantic information to be embedded in the shape of the signal. Study of sign change of P-operator on digital signals offered us several insights. One of them is peak and trough generation. Peaks and troughs contain most significant semantic information of a signal, because for a peak or a trough P-operator has to make a transition from negative to positive, which can be considered as a double jump in the ordered sequence $-, 0, +$. It has been proved in Lemma 3 that the positive to negative transition by P-operator is impossible and therefore all other transitions are only single jump in the sequence $-, 0, +$.

With the help of the features described in Theorem 3 (Table I) we have analyzed the raw signals, their shifted surrogates, PSD preserved shifted surrogates and Gaussian white noise signals and have shown their relative differences on two different datasets by simulation in Fig 3 and Fig 5. The same analysis can be carried forward to the signals and the noise that can contaminate them. In many instances it may be possible to effectively separate out the noise part from the signals with the help of the 13 features described in Theorem 3 along with appropriate machine learning algorithms.

It has been shown in Theorem 3 that 0 to 0 transition of P-operator is possible only in three ways. In one of them the signal must be strictly increasing, which can be associated with depolarization in an extracellular recording. In another of the three ways the signal must be monotonically decreasing, which can be associated with hyperpolarization in an extracellular recording. In a high sample frequency, high SNR extracellular signal distribution of these two features, along with the actual slope of the signal, may offer efficient neuronal spike sorting algorithms (see ref. [24] for a review).

Our next research direction will be to investigate into any possible relationship of the 13 features described in Table I of Theorem 3 with the underlying membrane and cellular level activities in intracellular and extracellular recordings. This is particularly significant, because we have shown in Theorem 3 that semantic information can be embedded in a digital signal only in 13 different 3-point configurations, all of which have been enumerated in Table I.

## Acknowledgment

This work was partly supported by a Department of Biotechnology, Government of India grant no. BT/PR7666/MED/30/936/2013.The authors like to thank Mr. Anupam Mitra for helping to understand the PSD preserved shifted surrogate signals.